\documentclass[aip,rsi,reprint,floatfix]{revtex4-1}
\usepackage{graphicx}
\usepackage{amsmath,amssymb}
\usepackage{color}
\usepackage[squaren,thinspace,textstyle]{SIunits}
\usepackage[normalem]{ulem}
\addunit{\evolt}{e\volt}
\newcommand{\er}{\ensuremath{\mu_\text{R}}}
\newcommand{\el}{\ensuremath{\mu_\text{L}}}

\newcommand{\vr}{\ensuremath{V_\text{gR}}}
\newcommand{\vl}{\ensuremath{V_\text{gL}}}

\newcommand{\algl}{\ensuremath{\alpha_\text{gL}^\text{L}}}
\newcommand{\algr}{\ensuremath{\alpha_\text{gR}^\text{L}}}
\newcommand{\argl}{\ensuremath{\alpha_\text{gL}^\text{R}}}
\newcommand{\argr}{\ensuremath{\alpha_\text{gR}^\text{R}}}

\newcommand{\sr}{\ensuremath{s_\text{R}}}
\renewcommand{\sl}{\ensuremath{s_\text{L}}}
\newcommand{\scr}{\ensuremath{s_\text{cr}}}

\newcommand{\vdqd}{\ensuremath{V_\text{DQD}}}

\begin{document}

\title{Determination of energy scales in few-electron double quantum dots}

\author{D.\ Taubert}

\affiliation{Center for NanoScience and Fakult\"at f\"ur Physik, Ludwig-Maximilians-Universit\"at M\"unchen, Geschwister-Scholl-Platz 1, D-80539 M\"unchen, Germany}
\author{D. Schuh}

\affiliation{Institut f\"ur Experimentelle Physik, Universit\"at Regensburg, D-93040 Regensburg, Germany}

\author{W.\ Wegscheider}

\affiliation{Laboratory for Solid State Physics, ETH Z\"urich, CH-8093 Z\"urich, Switzerland}

\author{S.\ Ludwig}

\affiliation{Center for NanoScience and Fakult\"at f\"ur Physik, Ludwig-Maximilians-Universit\"at M\"unchen, Geschwister-Scholl-Platz 1, D-80539 M\"unchen, Germany}

\begin{abstract}
The capacitive couplings between gate-defined quantum dots  and their gates vary considerably as a function of
applied gate voltages. The conversion between gate voltages and the relevant energy scales is usually performed in a
regime of rather symmetric dot-lead tunnel couplings strong enough to allow direct transport measurements.
Unfortunately this
standard
procedure fails for weak and possibly asymmetric tunnel couplings, often the case in realistic devices. We have
developed
methods to determine the gate voltage to energy conversion accurately in the different regimes of dot-lead
tunnel couplings and demonstrate strong variations of the conversion factors. Our concepts can easily be extended to
triple quantum dots or even larger arrays.
\end{abstract}


\maketitle

\section{introduction}\label{sec_introduction}

Electrostatically defined coupled quantum dot (QD) systems are interesting as an experimental toy model for fundamental
quantum-mechanical problems,\cite{heiblum,strongcoupling,eigenzitat1,twochannelkondo,backaction_old,eriksson,
zurichphonons,kataoka}
as qubits and registers for quantum information processing,
\cite{loss,hayashi,petta,koppens,pettacoherent,morello,wildmuell,silicon}
and to simulate molecular
electronics.\cite{taruchadots,spinpolarization}
In contrast to real molecules, the charge configuration and electronic spectrum of these artificial molecules are highly
tunable;\cite{tunable2,tunable1} the electronic energy scales thus have to be redetermined in each
experiment.
Usually the electronic spectra of coupled QDs are measured as a function of gate voltages. Hence, a
meaningful analysis of the measured data involves a conversion of the applied gate voltages to energy differences
between the electronic states. The conversion factors are specific to each sample, and even vary if the
configuration of a QD system is changed.

The conversion from gate voltages to energy can be achieved by comparison
with a known external energy scale. The most straightforward method relies on nonlinear transport
measurements where the external energy scale is provided by the applied source-drain voltage;\cite{dqd_review} this
method will be reviewed in Sec.\ \ref{sec_transport}. In the few-electron
regime,\cite{chargesensing2,fewelectron_petta} which is desirable for many applications, the
tunnel barriers of the double QD system are often too high to observe a current flow through the double QD.
Fortunately, charge fluctuations can still be measured, e.\,g., using a capacitively coupled quantum point contact (QPC)
as a charge detector.\cite{chargesensing2} With
this technique it is even possible to detect extremely small currents through an almost pinched-off double QD
indirectly by recording tunneling processes in real time.\cite{fcs} This counting method is
demanding, though, as it
needs a special setup, including a low-noise high-bandwidth detector. A more basic procedure which can be conducted
with a standard experimental setup would be desirable.

It is possible to extract energy scales from the thermal broadening of the transitions
between different charge configurations in the stability diagram of coupled QDs \cite{reconfline} in an elaborate
procedure. This method requires a small tunnel
coupling to guarantee that the line shapes are determined by thermal broadening. In an appropriate radio-frequency setup
and a double QD with suitable tunnel couplings photon-assisted
tunneling can be used for energy calibration since the photon energy provides the external energy scale.\cite{pat1,
dqd_review}

Most energy calibration methods including the aforementioned ones either require special experimental setups or a
narrow regime of tunnel couplings of the QDs. In practice the conversion factors are often
determined once in a rather open double QD by measuring nonlinear transport, and are then still used after tuning the
coupled QDs to rather small tunnel couplings.
In this article we show that such an approach leads to inaccuracies that can be avoided. We present
several
methods to acquire conversion factors, valid for different coupling regimes, which are accurate as well
as straightforward to implement. The discussion here is limited to double QD systems with rather weak tunnel coupling
between the two
QDs which, incidentally, have a very large resistance compared to the lead resistances; the latter can thus be
neglected.
Strong coupling between the QDs has been discussed in Ref.\ \onlinecite{strongcoupling}.

\section{sample}

\begin{figure}[h]
\includegraphics[width=\columnwidth]{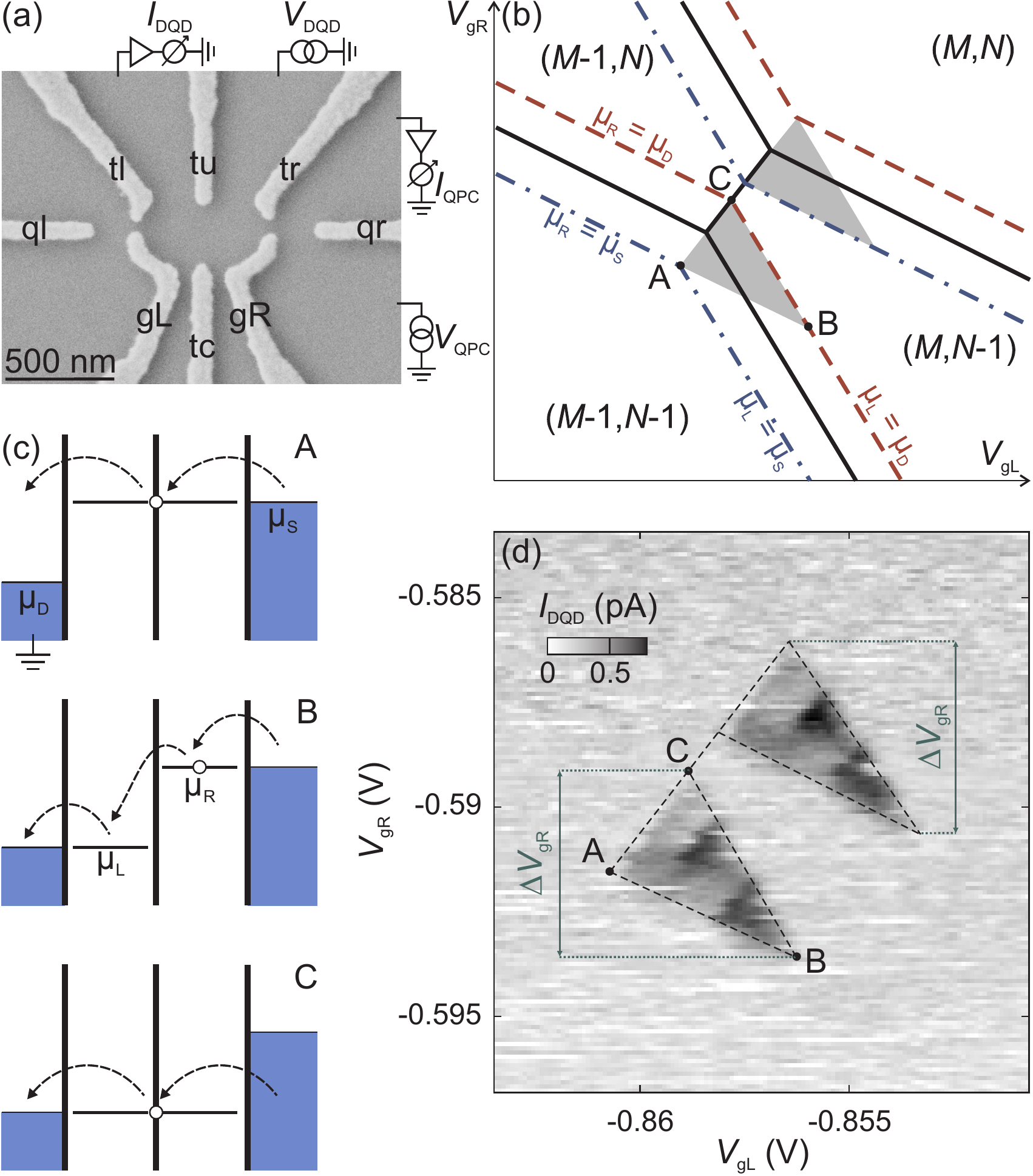}
\caption{\label{fig_sample}
(a) Scanning electron micrograph of a structure nominally identical to the actual sample. Eight gates (light gray) are
used to define a double QD as well as up to two QPC charge detectors. Electric circuits for driving current through the
double QD or the actually used QPC are indicated.
(b) Continuous (black) lines sketch a schematic charge stability diagram for an unbiased double QD. The splitting of
these charging lines, if a source-drain voltage \vdqd\ is applied across the double QD, into source and drain resonances
is depicted by interrupted (colored) lines (for details see main text). The gray triangles mark regions in which
first-order current is possible. The occupation numbers of the two QDs change from $(M-1,N-1)$ to $M,N$ from bottom left
to top right. Overall shifts of the diagram due to the applied bias are taken into account assuming equal capacitive couplings between the double QD and both leads (see Sec.\ref{capacitive} for details on capacitive coupling).
(c) Chemical potentials of the leads ($\mu_\text{S}$ for source and $\mu_\text{D}$ for drain) and QDs (\el, \er) at the positions marked by A, B, and C in (b). 
(d) Current through the double QD measured as a function of plunger gate voltages \vl\ and \vr\ for $\vdqd = -241.1
\mu$V. Regions of finite current are enclosed by dashed lines.
}
\end{figure}

The sample has been fabricated from a GaAs/AlGaAs heterostructure containing a two-dimensional electron system (2DES)
\unit{85}{\nano\meter} below the surface. The charge carrier density of the 2DES was $n_s= \unit{1.9 $\times$
\power{10}{11}} {\mathrm{cm}^{-2}}$ and
its mobility was 
$\mu = \unit{1.2 $\times$ \power{10}{6}}{\centi\meter\squared\per(\volt\second)}$ at low temperatures. The
measurements presented here have been performed at an electron temperature of $T\simeq50$\,mK.
The double QD is defined by applying negative voltages to metallic electrodes on the sample surface which have
been created by electron beam lithography. A scanning electron micrograph of a nominally identical structure is shown
in Fig.\ \ref{fig_sample}(a). 
The design is similar to the one presented in Ref.\ \onlinecite{design}. The plunger gates ``gL'' and ``gR'' are used
to change the chemical potentials of the two QDs L and R while the tunneling barriers between the QDs as
well as between each QD and its adjacent lead are controlled by four additional gates ``tu,'' ``tc,'' ``tl,'' and
``tr'' (these
gates also influence the chemical potentials but are kept constant during a measurement). Gate ``qr'' together
with
tr and gR defines the QPC that is used as charge detector. A second QPC can be defined on 
the left-hand side of the sample. The voltage $V_\text{qr}$ is varied proportionally to the two
plunger gate voltages to compensate for the capacitive coupling between plunger gates and the QPC and to keep the QPC
current roughly constant during a measurement.

\section{basic relations}

We define the chemical potential $\mu_\text{QD} (N)$ of a QD occupied by $N-1$ electrons as the energy needed to add the
$N$th electron. If only a very small bias voltage is applied to the double QD, the first-order linear-response current,
$I_\text{DQD}$, through the double QD only flows at the triple points where the
chemical potentials of both QDs ($\mu_\text{L}$ and $\mu_\text{R}$) and the leads ($\mu_\text{S}$ and
$\mu_\text{D}$) are aligned ($\mu_\text{L} = \mu_\text{R} = \mu_\text{S} = \mu_\text{D}$). A charge stability diagram
containing four different charge configurations ($M,N$) is shown schematically in Fig.\ \ref{fig_sample}(b) in black
(continuous) lines for negligible bias voltage. Along these charging lines
the chemical potential of one QD is in resonance with the Fermi energy of  the leads ($\mu_{\rm R} = \mu_\text{S} =
\mu_\text{D}$ or $\mu_{\rm L} = \mu_\text{S} = \mu_\text{D}$).

If a finite bias is applied, resonances with
the source and the drain lead occur for different plunger gate voltages. This is also shown exemplarily in Fig.\
\ref{fig_sample}(b) where we choose $\mu_\text{S} > \mu_\text{D}$. The source resonances $\mu_{\rm L,R}=\mu_\text{S}$
are plotted as blue (dashed-dotted) lines and the drain resonances $\mu_{\rm L,R}=\mu_\text{D}$ as red (dashed) lines.
The two triple points of finite current (by first-order sequential single-electron tunneling) grow into two identical,
so-called ``bias triangles'' with $\mu_\text{D} \le \el \le \er \le \mu_\text{S}$, which are marked
in gray in Fig.\ \ref{fig_sample}(b).\cite{dqd_review} The left triangle in Fig.\ \ref{fig_sample}(b) corresponds to the addition of the first of two electrons. For its three corners, A, B, and C, the alignment of the chemical potentials is sketched in Fig.\ \ref{fig_sample}(c). Note that the position of the source and drain resonances in a stability diagram [such as the schematic in Fig.\ \ref{fig_sample}(b)] depend directly on $\mu_\text{S}$ and $\mu_\text{D}$ as well as on the capacitive couplings between the QDs and the leads (see Sec.\ \ref{capacitive} for details).

The chemical potentials $\mu_\text{QD}$ of the QDs are controlled by the voltages $V_{\text{g}i}$ on plunger gates g$i$
via their capacitive coupling expressed by the conversion factor $\alpha_{\text{g}i}^\text{QD} \equiv
- \frac{\partial \mu_\text{QD}}{\partial V_{\text{g}i}}$. Considering two plunger gates and a double QD, we find 
\begin{eqnarray}
d\el &= -\algl d\vl - \algr d\vr\,,\nonumber\\
d\er &= -\argl d\vl - \argr d\vr\,.\label{eq_leverarms}
\end{eqnarray}
From this the slopes $s_\text{QD}=d\vr / d\vl$ of the charging lines parallel to AB (CB) in Fig.\
\ref{fig_sample}(b), 
\begin{eqnarray}
\sr &=& - \frac{\argl}{\argr}\,;\ d\er=0\,,\nonumber\\
\sl &=& - \frac{\algl}{\algr}\,;\ d\el=0\,,\label{eq_srsl}
\end{eqnarray}
as well as the slope of the charge reconfiguration line along AC,
\begin{eqnarray}
\scr &=& \frac{\algl - \argl}{\argr - \algr}\,;\ d\er=d\el \label{eq_scr}\,,
\end{eqnarray}
can be easily derived. In a sufficiently small region of the stability diagram all these slopes and conversion
factors are constant, which is equivalent to constant capacitive couplings. Under this condition ABC in Fig.\
\ref{fig_sample}(b) forms a triangle and Eqs.\ (\ref{eq_leverarms})--(\ref{eq_scr}) allow an accurate coordinate
transformation from plunger gate voltages to energy values of the QDs' chemical potentials.

\section{Calibration in nonlinear transport measurements}
\label{sec_transport}

In the simplest case the bias triangles sketched in Fig.\ \ref{fig_sample}(b) can be measured directly. Such an example
is depicted in Fig.\ \ref{fig_sample}(d) which shows the current flowing through the double QD for a bias of $\vdqd =
-241.1 \mu$V applied to the source lead. Nonzero current is observed within two almost identical triangles just as
sketched in Fig.\ \ref{fig_sample}(b). The conversion factors can be extracted  from the dimensions of these triangles.
We can start, e.g., by measuring the voltage change, $\Delta\vr$, between point B and point C along the  $\vr$ axis in 
the stability diagram of Fig.\ \ref{fig_sample}(d). For the transition B $\rightarrow$ C, we find $\Delta\el=0$ and
$\Delta\er=e\vdqd$ [compare Fig.\ \ref{fig_sample}(c)]. These relations and equations
(\ref{eq_leverarms})--(\ref{eq_scr}) allow us to determine all
relevant conversion factors,
\begin{equation}\begin{array}{ll}\label{eq_alphas}
\argr = \frac{\left| e \vdqd \right|}{\Delta \vr} \frac{\sl}{\sl-\sr}\,,\quad
&\argl = - \sr \argr\,,\\\\
\algr = \frac{\scr -\sr}{\scr - \sl} \argr \,,
&\algl = - \sl \algr\,.
\end{array}\end{equation}
The numerical values obtained from the current measurements in Fig.\ \ref{fig_sample}(d) are
shown in Table \ref{tab_results}. Error values are also given, which reflect the limited accuracy of determining the slopes and voltage differences from the greyscale plot. Averaging over both current-carrying triangles has been used where applicable, with the error value representing the standard deviation. The conversion factors presented in Table \ref{tab_results} have to be understood as a lower limit of the actual conversion factors due to the limited sensitivity of the current measurement, though, as will be discussed in detail in Sec.\ \ref{sec_results}. The error values do not include this systematic error.

\begin{table*}
 \begin{tabular}{lcccc}\hline\hline
 & Fig.\ \ref{fig_sample} & Fig.\ \ref{fig_triangle} & Fig.\ \ref{fig_kink} & Fig.\ \ref{fig_backaction}\\\hline
\sr 			& $-0.48 \pm 0.03$  	& $-0.51 \pm 0.03$ 	& $-0.60 \pm 0.04$	&$-0.58 \pm 0.03$ \\
\sl  			& $-1.59 \pm 0.18$ 	& $-1.71 \pm 0.03$ 	& $-1.81 \pm 0.09$	& $-1.77 \pm 0.01$\\
\scr 			& $1.29 \pm 0.10$ 	& $1.48 \pm 0.12$ 	&$0.98 \pm 0.08$	& $0.70 \pm 0.07$ \\\hline
$\Delta \vr$ (mV) 	& $4.52 \pm 0.08$	& $4.22 \pm 0.12$ 	& $3.86 \pm 0.06$	&$5.61\pm 0.11$\\
\argl\ (meV/V) 		& $37 \pm 6$		& $41 \pm 6$ 		&$56 \pm 7$		& $60 \pm 6$\\
\argr\ (meV/V) 		& $76 \pm 8$ 		& $81 \pm 6$ 		&$94 \pm 7$		& $103 \pm 5$\\
\algr\ (meV/V) 		& $47 \pm 8$		& $51 \pm 6$ 		&$53 \pm 7$		&$ 53 \pm 4$\\
\algl\ (meV/V) 		& $75 \pm 5$ 		& $86 \pm 7$ 		& $96 \pm 7$		&$94 \pm 7$\\\hline
\vdqd\ ($\mu$V) & -241.1 & -241.1 & -241.1 & -387.5 \\
$V_\text{tl}$ (V) & -0.415& -0.410 & -0.435 & -0.380\\
$V_\text{tr}$ (V) & -0.430& -0.425& -0.450 & -0.465\\\hline\hline
 \end{tabular}
\caption{\label{tab_results} Summary of the charging line slopes and conversion factors determined from the presented data
for the four different double QD configurations shown in Figs.\ \ref{fig_sample}--\ref{fig_backaction}. \vdqd\ as well
as gate voltages that are changed between measurements are also noted for convenience.}
\end{table*}

\section{Calibration in charge measurements}
\label{sec_charge}
\begin{figure}[h]
\includegraphics[width=\columnwidth]{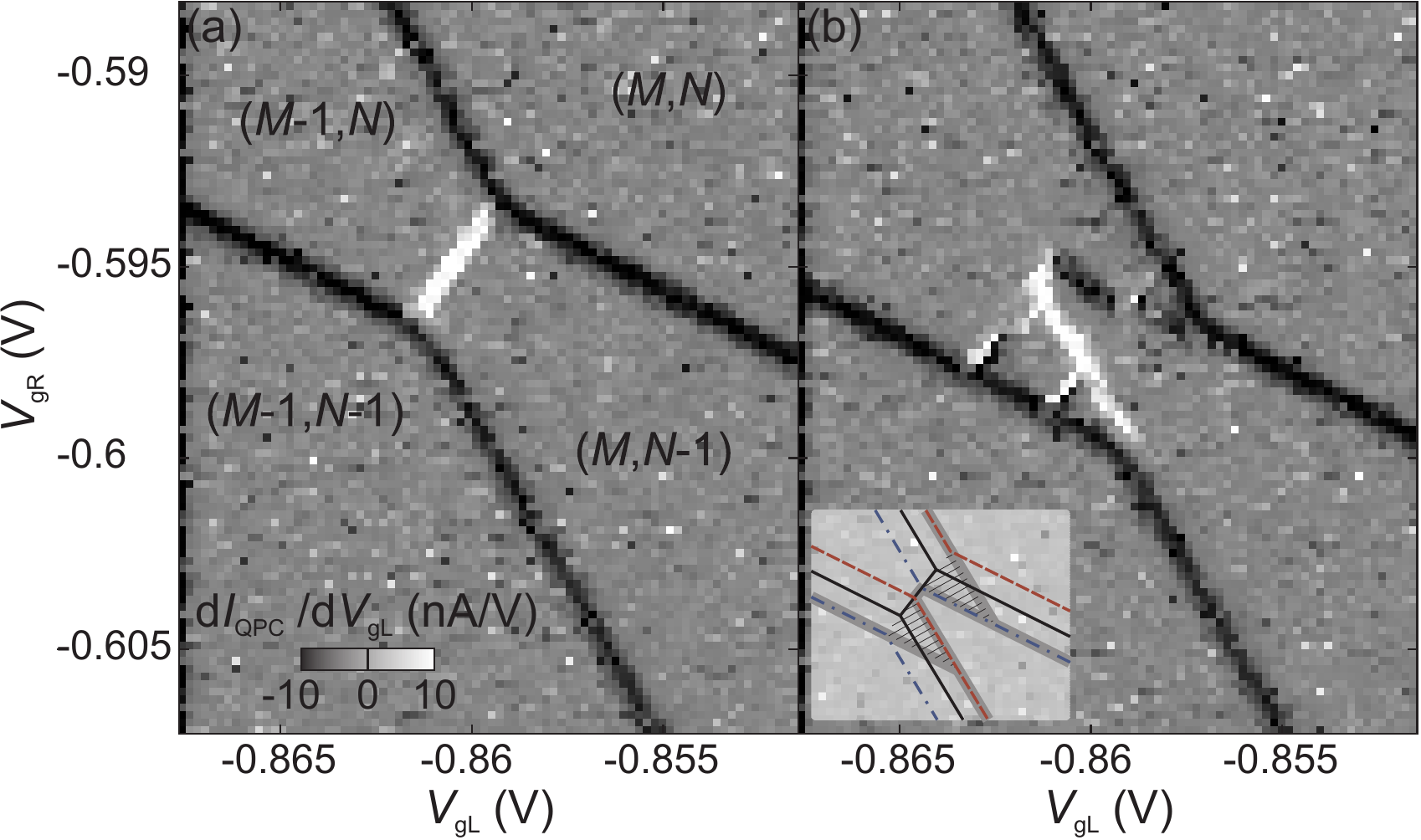}
\caption{\label{fig_triangle}
(a) Charge stability diagram (compare Fig.\ \ref{fig_sample}) of the double QD measured with the right QPC used as
charge detector. The transconductance $\text{d}I_\text{QPC}/\text{d} \vl$ (determined by taking the numerical derivate of
the dc current $I_\text{QPC}$) is plotted for (a) $\vdqd\simeq0$ and (b) $\vdqd = -241.1 \,\mu$V. Applied bias and gate
voltages are very similar as in Fig.\ \ref{fig_sample}(d). Inset: Schematic charge stability diagram similar to Fig.\
\ref{fig_sample}(b). Charging lines actually observed in (b) are marked in gray, the bias triangles that would be
visible in nonlinear transport are shaded.}
\end{figure}

Instead of measuring current through the double QD, recording its charge by means of a QPC charge
detector allows a determination of the conversion factors with a higher accuracy. In a measurement with zero bias
applied to the double QD [see Fig.\ \ref{fig_triangle}(a)] 
the charging lines (dark) and charge reconfiguration lines (white) mark the boundaries between stable charge
configurations. For rather symmetric tunnel couplings to both leads the
triangles observed for nonlinear transport may show up again as regions of constant average charge
\cite{spin_blockade_triangles} if a bias voltage is applied to the double QD. Such a situation is depicted in
Fig.\ \ref{fig_triangle}(b).  At finite bias, the white
charge reconfiguration line is only visible along the short section connecting the bases of the two triangles.
Additional parallel lines can be attributed to excited states in one of the QDs.\cite{excited_states}

The conversion factors are extracted in the same way as for
Fig.\ \ref{fig_sample}, using Eq.\ (\ref{eq_alphas}) and  again averaging over both triangles. It is
helpful to take the slope of the charge reconfiguration line from the measurement with zero applied bias to increase
the accuracy. The resulting slopes and conversion factors can be found in Table \ref{tab_results}.

\begin{figure}[h]
\includegraphics[width=\columnwidth]{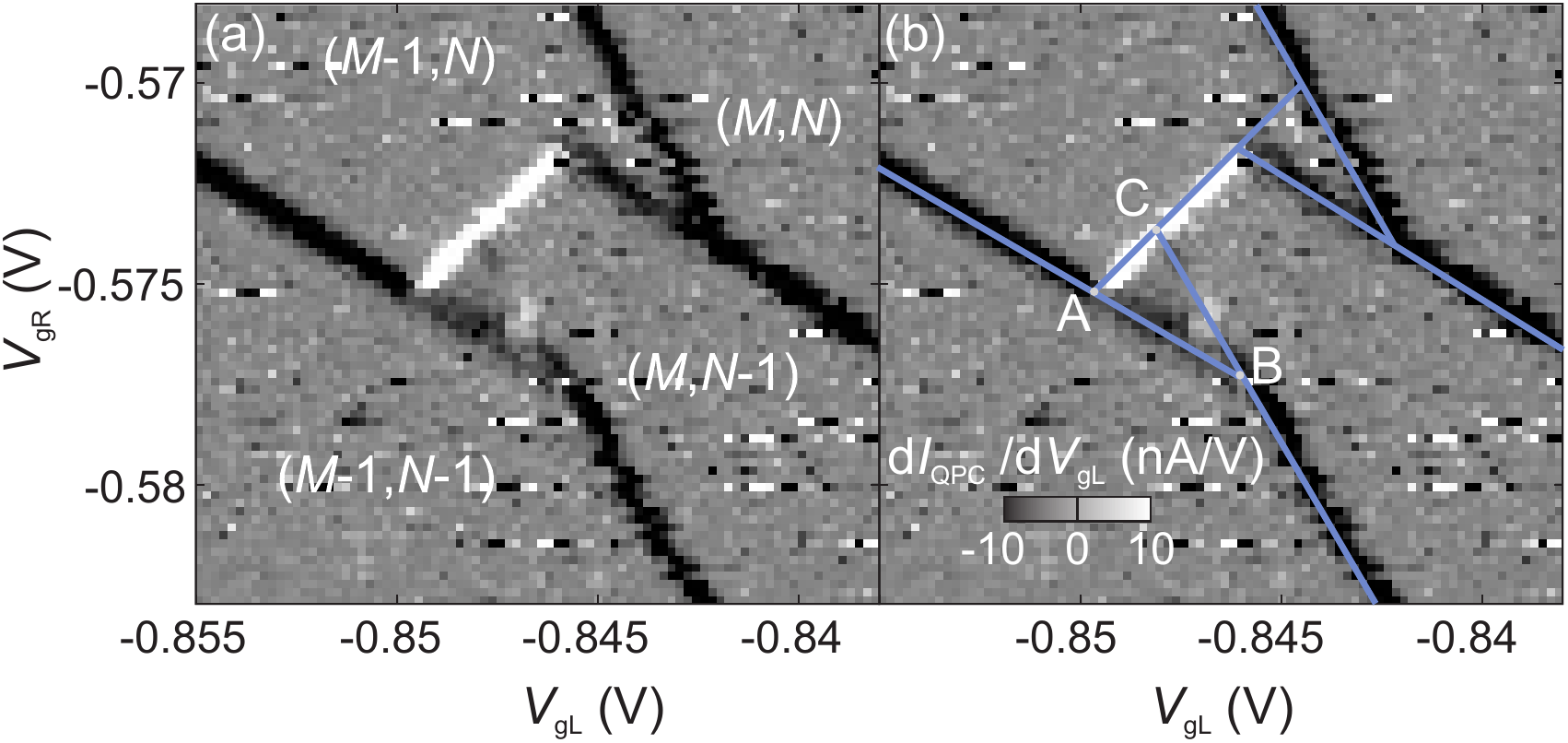}
\caption{\label{fig_kink} 
(a) Charge stability diagram as in Fig.\ \ref{fig_triangle} ($\vdqd =-241.1\,\mu$V) but with overall reduced and
asymmetric tunnel couplings. The tunnel barrier between the double QD and the drain lead is larger than the other tunnel
barriers. Distinct kinks
in the black charging lines are reminiscent of the bias triangles and can be used for determining the conversion factors.
(b) Same data as in (a); reconstruction of the bias triangles (solid lines).
}
\end{figure}

If at least one of the three relevant tunnel barriers of the double QD (the QD-lead barriers or the interdot
barrier) is increased, $I_\text{DQD}$ will decrease, eventually becoming too small to be measured. In this
case the energy calibration \emph{has to be} performed via charge detection. The features that appear in such a
nonlinear stability diagram depend on the ratio of the three relevant tunnel couplings. For instance, the bias
triangles are only observed for rather symmetric
tunnel couplings.
Fig.\ \ref{fig_kink}(a)
depicts a typical situation
in which the tunnel coupling to the drain lead is small (due to a high barrier) compared to the other two relevant
tunnel couplings. As a
result the drain
resonance of the left bias triangle is not observed and the average charge configuration within this bias triangle is
close to $(M, N-1)$. The charge reconfiguration line (white) is therefore clearly visible in Fig.\ \ref{fig_kink}(a).
From the kink in the charging lines, the bias triangles can still be reconstructed as demonstrated in Fig.\
\ref{fig_kink}(b), allowing the usage of the same calibration relations as described above.
The results have been added to Table \ref{tab_results}.

\section{asymmetrical configuration}
\label{sec_backaction}

Finally we study an even more asymmetric  double QD system, with the barrier between the right QD and the
adjacent source lead almost closed. Such a situation is favorable for experiments studying, e.g., backaction of the QPC
on the
double QD\cite{backaction_old,backaction_daniel}. Current through the double QD is almost completely blocked
($I_\text{DQD}\simeq0$).

\begin{figure}[h]
\includegraphics[width=\columnwidth]{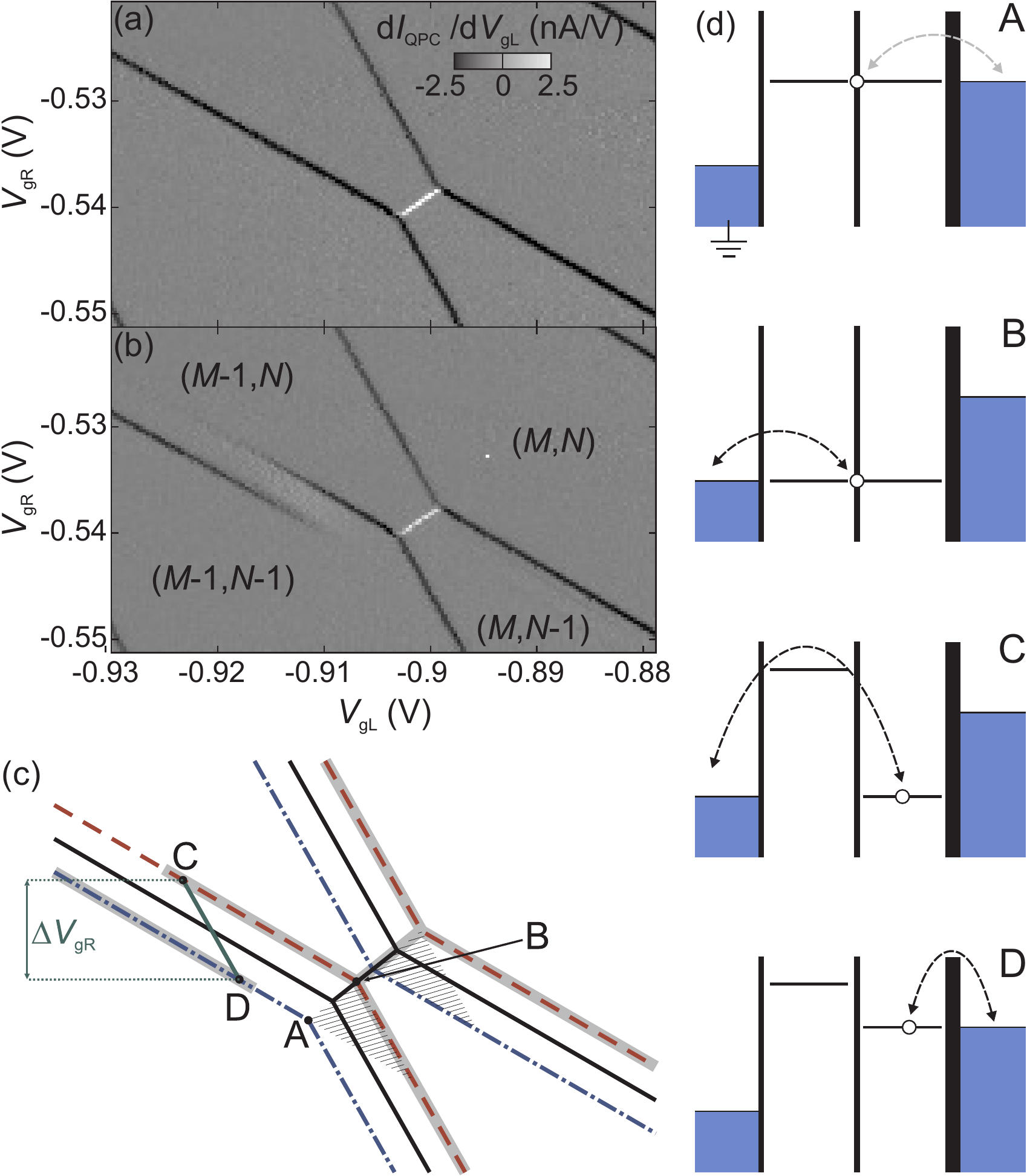}
\caption{\label{fig_backaction} 
(a) Charge stability diagram of a double QD with very asymmetric tunnel couplings at $\vdqd\simeq0$. The tunnel barrier
between double QD and source lead is almost closed.
(b) Measurement as in (a) but for $\vdqd=-387.5\,\mu$V.
(c) Schematic charge stability diagram as in Fig.\ \ref{fig_sample}(b). Charging lines are underlined in gray where visible in (b). The voltage difference $\Delta\vr$ between positions C and D can be used for determining the conversion factors (for details see main text).
(d) Chemical potentials for the positions marked A, B, C, D in (c). The source resonance in ``A'' is not observed in
(b), since charge exchange through the very large source barrier (denoted by the thick black line in the level diagrams)
is almost blocked.
}
\end{figure}

Fig.\ \ref{fig_backaction}(a)
shows the charge stability diagram of such a double QD system with zero bias voltage in comparison to Fig.\
\ref{fig_backaction}(b) where a large negative bias is applied to the right-hand side (source)
lead. No indications of triangles can be observed in Fig.\ \ref{fig_backaction}(b), though. Most of the charging lines
visible in Fig.\
\ref{fig_backaction}(a) are resonances with
the left (drain) lead since the left tunnel barrier is reasonably low. In the vicinity of the triple points
($\el\simeq\er$) charging of the right QD occurs from the left (drain) lead in an elastic cotunneling process
via the left QD. The charging line belonging to the transition $(M-1,N-1) \leftrightarrow (M-1,N)$ is
discontinuous, however, as further away from the triple point the cotunneling rate quickly decreases and the
right QD can be charged more efficiently from the right-hand side (source) lead via the large tunnel barrier.
The charging line of the right QD therefore shows a step as the transition from drain to source resonance occurs, 
with a step size determined by $e \vdqd$. In Fig.\ \ref{fig_backaction}(c) a sketch of the charge stability diagram
including source and drain resonances is shown, with the lines visible in Fig.\ \ref{fig_backaction}(b) marked in gray.
 The chemical potential of the double QD system
at the points A, B, C, and D are depicted in Fig.\ \ref{fig_backaction}(d). The points C and D have been chosen such that the line CD is parallel to the charging
line of the left QD. At C, the right QD is in resonance with the left lead whereas at D the right lead is resonant, so $\Delta \el = 0$ and $\Delta \er = e \vdqd$ between C and D, just as between B and C in Figs.\ \ref{fig_sample}--\ref{fig_kink}. Hence, the distance CD determines $\Delta \vr$. To obtain the conversion
factors listed in Table \ref{tab_results} from Eq.\ (\ref{eq_alphas}) we additionally use the slopes of the charging
lines, \sl, \sr, and \scr.

\section{comparison of results}
\label{sec_results}
The conversion factors listed in Table \ref{tab_results} vary considerably between the four measurements, even
though the data shown in Figs.\ \ref{fig_sample} -- \ref{fig_backaction} have all been measured within the same
cooling run on the same sample. The main difference between those measurements is that the double QD is tuned to
slightly different geometries by varying the gate voltages
$V_\text{tl}$ and $V_\text{tr}$ as shown in Table \ref{tab_results}
($V_\text{tu} = \unit{-0.440}{\volt}$ and $V_\text{tc} = \unit{-0.310}{\volt}$ are kept constant).
For instance the conversion factors of Fig.\ \ref{fig_backaction} are up to 62 \% higher than those of Fig.\
\ref{fig_sample}. We conclude that the capacitances between gates and the QDs depend on the detailed QD geometries
which are strongly affected by the applied gate voltages. Whenever accurate energy values of the electronic states in
coupled QDs are desired, it is therefore of utmost importance to perform a calibration right at the gate voltage
settings of interest. In this article we demonstrate that simple calibration methods are available for very different
coupling regimes.

The voltage settings for Figs.\ \ref{fig_sample} and \ref{fig_triangle} are almost identical, and the same should be
true for the conversion factors. Nevertheless the $\alpha$'s listed in Table
\ref{tab_results} are up to 14 \% higher for Fig.\ \ref{fig_triangle} than those determined from Fig.\ \ref{fig_sample}.
This discrepancy is caused by the difficulty and according inaccuracy in determining the size of the current-carrying
triangle in Fig.\ \ref{fig_sample}, since the observed size depends on the sensitivity of the current measurement. This causes a systematic error in $\Delta\vr$ which is not included in the error value given in Table \ref{tab_results}.

In addition, in a current measurement the size of the bias triangle is influenced by the width of the transitions between
the different charge configurations (i.e., the width of the charging lines in a charge measurements) which depends on
the tunnel couplings. The accuracy is much higher for charge detection, where $\Delta\vr$ can be determined by using the center of the charging lines independent of their width. Hence, we recommend
using charge detection or, at least, the derivative $dI_\text{DQD}/dV_{\text{g}i}$ of the current for energy
calibration.

\section{Interplay of the different methods}

\begin{figure}[h]
\includegraphics[width=\columnwidth]{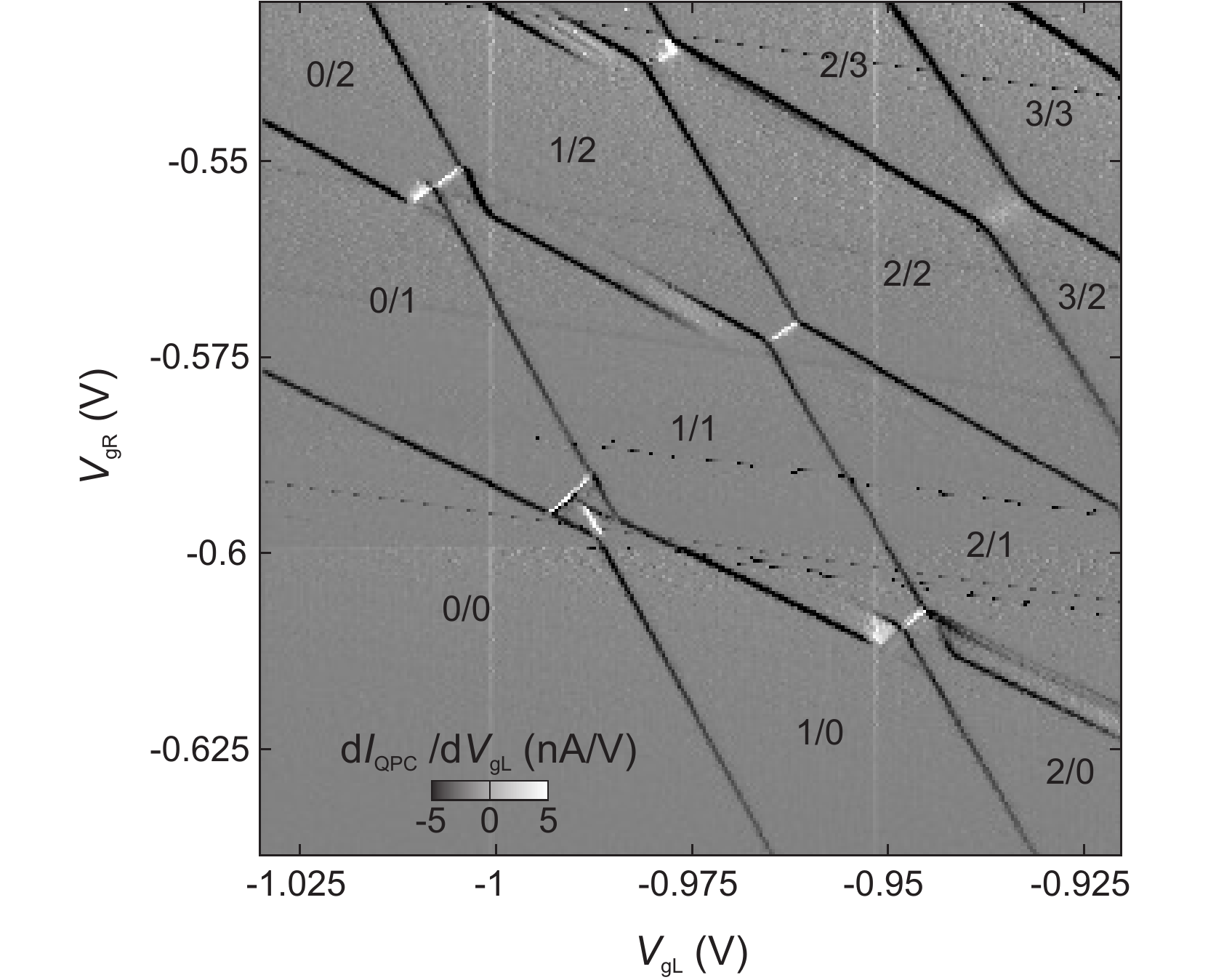}
\caption{\label{fig_overview} 
Larger section of the charge stability diagram for an asymmetrically tunnel coupled double QD similar to that of
Fig.\ \ref{fig_backaction}(b) at $\vdqd=-387.5 \,\mu$V. Absolute values of the electron numbers in the left ($L$) and
right QD ($R$) are given as $L/R$. Barely visible almost horizontal lines are related to resonances in the detector QPC.
}
\end{figure}

In Fig.\ \ref{fig_overview} 
a larger portion of a charge stability diagram (measured using charge detection) is shown which features all of the
effects
described in Sections \ref{sec_charge}--\ref{sec_backaction} as the ratio between the relevant tunnel couplings varies
throughout the stability diagram as a function of the plunger gate voltages. For this measurement the double QD has
been tuned rather
asymmetrically, similar to the configuration in Fig.\ \ref{fig_backaction}, and the feature visible in the latter graph
reoccurs here as the electron numbers change from 1/1 to 2/2. As the tunnel couplings vary, different effects are
observed at other transitions. The feature visible at the transition 0/0 $\leftrightarrow$ 1/1 is very
similar to that of Fig.\ \ref{fig_triangle} while the transition 0/1 $\leftrightarrow$ 1/2 resembles Fig.\ \ref{fig_kink}.
The features visible at  1/0 $\leftrightarrow$ 2/1 contains mixed signatures. The region 2/2 $\leftrightarrow$ 3/3 shows
no signs of $\vdqd\ne0$. Here our calibration methods fail and the conversion factors have to be estimated by
extrapolation from other areas of the stability diagram. Calibration methods which need specific experimental setups
might still work (compare Sec.\ \ref{sec_introduction}).

\section{Capacitive coupling between quantum dots and leads}
\label{capacitive}
\begin{figure}[h]
\includegraphics[width=\columnwidth]{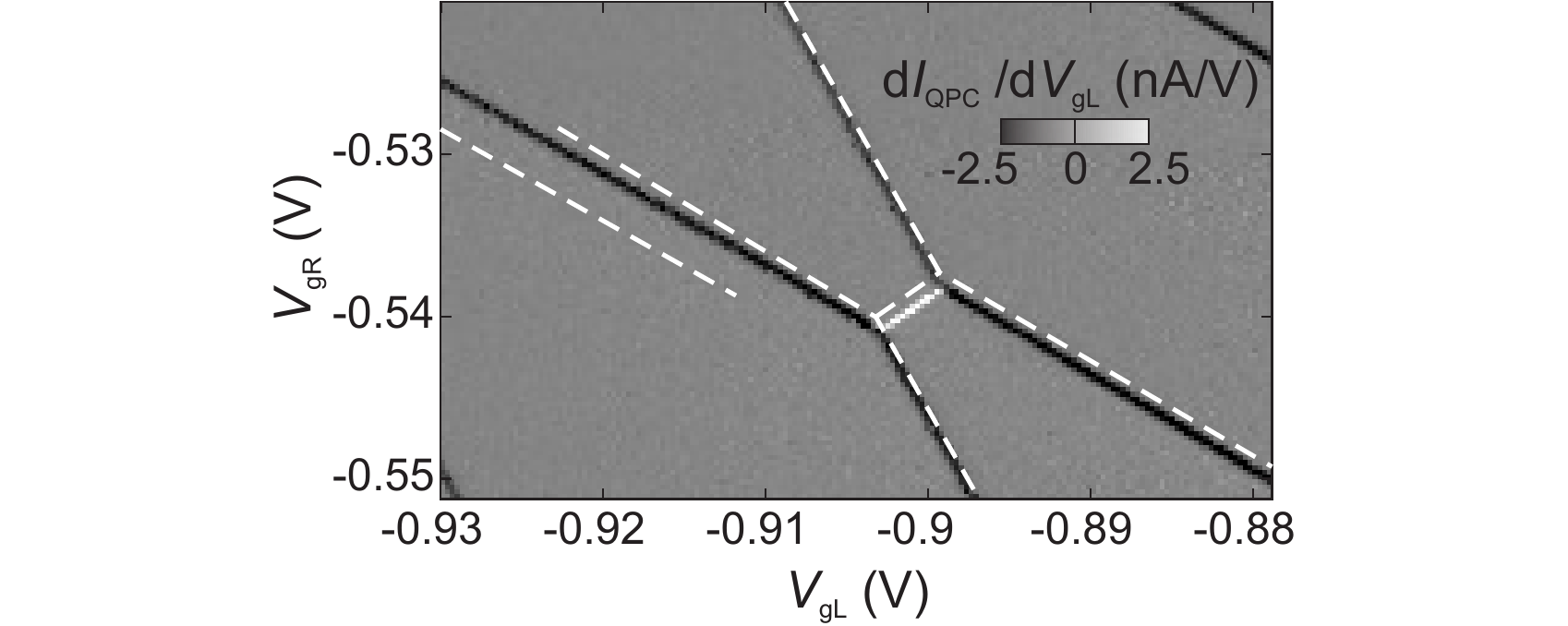}
\caption{\label{fig_shift} The gray-scale plot shows the zero-bias data from Fig.\ \ref{fig_backaction}(a), with the
charging lines from Fig.\ \ref{fig_backaction}(b) ($\vdqd = -387.5 \,\mu$V) superimposed as dashed lines.}
\end{figure}

All calibration methods described so far rely on distinct features that appear in the stability diagram as a finite
but fixed bias \vdqd\ is applied. It is tempting to utilize instead the linear response of the position of a
charging line as \vdqd\ is varied. Fig.\ \ref{fig_shift}
illustrates the overall effect of applying a bias voltage. The zero-bias data from Fig.\ \ref{fig_backaction}(a) are reproduced, with the charging lines from the $\vdqd=-387.5\,\mu$ data in Fig.\ \ref{fig_backaction}(b) superimposed (dashed lines).
Changing the bias results in a much smaller shift of the charging lines than expected from the split charging line on
the left-hand side that has been used for calibration in Sec.\ \ref{sec_backaction}. The observed shift is therefore not
a direct measure of  $e\vdqd$. It is much smaller than expected from the applied bias \vdqd\ because of a compensation
related to the electrostatic coupling between the QDs, the gates, and the leads. 
As pointed out with this example, relying solely on changes in the charging line
positions is  not suitable to obtain a valid calibration. Instead, we used features that show transitions from source to
drain resonances.

\section{conclusions}

We have described several methods of obtaining the conversion factors between gate voltages and chemical
potentials in a double QD system in the few-electron regime. Depending on the specific couplings and
tunneling rates in the system, different phenomena are observed in the charge stability diagram as a bias is applied
across the double QD, and most of them can be harnessed to determine the conversion factors. Even for a very asymmetric
system, a calibration procedure has been developed. The latter method might gain significance if the number of quantum
dot in series is increased, as additional QDs tend to act in the same way as large barriers when in Coulomb blockade.
The procedure might therefore be extended to few-electron triple QD circuits which include a charge
detector\cite{tripledot_munich, tripledot_hannover, tripledot_canada, granger}.

\begin{acknowledgments}

We thank D.\ Harbusch for his meticulous improvements of the cryostat's cabling and filtering. Financial
support by the German Science Foundation via SFB 631, LU 819/4-1, and the German
Excellence Initiative via the "Nanosystems Initiative Munich (NIM)" is gratefully
acknowledged. 
\end{acknowledgments} 


%
%
\end{document}